\documentclass[twocolumn]{aastex701}

\usepackage{amsmath}
\usepackage{hyperref}
\usepackage{cleveref}
\defcitealias{adams2020energy}{A20}

\begin{document}

\title{Intra-system Uniformity through Planetary Embryo Accumulation}

\author[0009-0003-3584-6698]{Donald Liveoak}
\affiliation{Department of Physics, University of Michigan, Ann Arbor MI 48109, USA}
\email{dliveoak@umich.edu}

\author[0000-0002-8167-1767]{Fred C. Adams}
\affiliation{Department of Physics, University of Michigan, Ann Arbor MI 48109, USA}
\affiliation{Department of Astronomy, University of Michigan, Ann Arbor MI 48109, USA}
\email{fca@umich.edu}

\correspondingauthor{Donald Liveoak}
\email{dliveoak@umich.edu}

\begin{abstract}
Intra-system uniformity is a key trend in observed exoplanet systems. Understanding the physical mechanisms which sculpt planetary systems into uniform or non-uniform configurations will help constrain theories of planet formation. Motivated by previous work showing that intra-system uniformity is a natural consequence of proto-planetary systems dissipating energy and settling into lower energy configurations, this paper explores the energy minimization hypothesis using $N$-body simulations. We find that the accumulation of planetary embryos through inelastic mergers generally dissipates a substantial fraction of the system energy, but the systems do not always reach the global energy minimum. Analytic predictions indicate that planetary pairs have nearly equal masses in their lowest energy state when the total mass falls below a mass threshold. With larger total mass, one member of the pair tends to accrete most of the mass. Our numerical simulations show that pairs with masses above the threshold tend towards non-uniformity, but that the full realization of the effect occurs at a larger mass scale, a factor of $\sim4$ above the threshold. In any case, however, the low-mass planet pairs are generally more uniform than pairs with high mass, in agreement with previous work. Finally, we compare our results to the observed sample of exoplanets and find overall agreement.
\end{abstract}


\section{Introduction}
Over recent decades, thousands of exoplanets have been discovered, broadening our understanding of the mechanisms which sculpt planetary systems. Planetary systems in the observational sample often consist of several low-mass planets in compact configurations \citep{lissauer2011architecture}. These \textit{compact multis} tend to exhibit a high degree of intra-system (often called ``peas-in-a-pod") uniformity in mass \citep{millholland2017kepler}, radius \citep{weiss2018california,weiss2018california2}, and period ratio \citep{weiss2018california}. Additionally, the orbits of compact multis tend to be approximately circular and coplanar \citep{fang2012architecture}.

Despite suggestions that these trends may arise from observational biases \citep{murchikova2020peas}, intra-system uniformity has been shown to be statistically robust \citep{he2019architectures} and astrophysical in nature \citep{weiss2020kepler}. Elucidating the physical mechanism(s) responsible for intra-system uniformity (or the lack thereof) among observed planetary systems allows us to better understand and constrain theories of planet formation.

Previous studies have explored intra-system uniformity through population synthesis models \citep{mishra2021new, emsenhuber2023planetary}, $N$-body simulations \citep{goldberg2022architectures, lammers2023intra, ghosh2024orbital}, and analytical approaches \citep{tremaine2015statistical}.

In addition, intra-system uniformity can be understood as a natural outcome of a general class of formation scenarios in which energy dissipation is substantial and angular momentum is conserved \citep{adams2019pairwise}.
Specifically, the orbital energy of two adjacent forming planets is minimized when both embryos acquire nearly equal mass and attain circular and coplanar orbits. Subsequent work (\cite{adams2020energy}, hereafter \citetalias{adams2020energy}) extended this result to account for the gravitational self-energy of the forming planets and found that when the total mass of neighboring bodies exceeds a certain critical value, it becomes energetically favorable for one planet to dominate and accrete most of the available material. This result naturally explains the observed trend that intra-system uniformity is disrupted for systems containing giant planets \citep{wang2017rv}. Since the mass threshold decreases with orbital radius, this scenario also predicts that the observed uniformity cannot extend beyond a few AU. 

The analysis of \citetalias{adams2020energy} applies to a variety of formation scenarios but makes two key assumptions. First, in order for intra-system uniformity to be the minimum energy configuration of the system, the energy minimization must be carried out in a pairwise manner, so that each forming planet is only subject to the gravitational influence of its immediate neighbor(s). Secondly, the mechanism of energy dissipation must be rapid enough and persist for long enough to allow the system to settle into a near-optimal configuration.

In this paper, we explore the energy minimization hypothesis in the context of pairwise planetary embryo accumulation using $N$-body simulations. We model the late stages of planet formation as a series of inelastic mergers of planetary embryos of randomized mass which dissipate orbital energy and gravitational self-energy. For this study, we do not model the process of pebble accretion, which serves as an alternative channel for energy dissipation \citep{lambrechts2014separating, bitsch2015growth}. Instead, we explore the extent to which pairwise collisions drive the systems to mass uniformity.

We find that the energy dissipation in our model causes systems to approach the minimum energy configuration derived in \citetalias{adams2020energy}, although it is not rapid enough to consistently reach the global energy minimum. Furthermore, we analyze the observed exoplanet sample and find that planet pairs above the critical threshold derived in \citetalias{adams2020energy} tend to exhibit less mass uniformity.

The remainder of this paper is organized as follows. In Section \ref{sec:methods}, we detail our numerical set-up and highlight our model assumptions. In Section \ref{sec:results}, we analyze the resulting intra-system uniformity of our simulated systems, demonstrating overall agreement with \citetalias{adams2020energy}. In Section \ref{sec:discussion}, we evaluate our model's assumptions and compare our results to the observed exoplanet sample. We conclude in Section \ref{sec:conclusion} with a summary of our results and a brief discussion of their implications. 

\section{Methods}\label{sec:methods}

To evaluate energy minimization as a mechanism for generating intra-system uniformity, we consider the late stages of planet formation, where sub-Earth mass planetary embryos accumulate to form planets. Here we model the protoplanetary system as planetary embryos whose masses are a few percent of $M_\oplus$, corresponding to the later stages of the planet formation process. For typical disk masses $\sim 0.5\times 10^{-4}-10^{-3}\, M_\star$ \citep{hartmann2008masses, mann2009circumstellar, pascucci2016steeper}, this choice corresponds to $N_e\sim 10^3-10^4$ embryos, which is on the upper end of what is computationally feasible given that simulation run-time scales as $\mathcal{O}(N_e^2)$ \citep{rein2012rebound}.

We consider distributions of planetary embryos which jointly comprise a protoplanetary disk with a power-law surface density $\Sigma(r) \sim r^{-p}$, where $p$ is typically inferred to be $\sim 1-2$ in planet formation scenarios (see \citealt{weiss2022architectures} and references therein). We truncate the disk at minimum and maximum orbital distances $r_{\min}$ and $r_\text{max}$ so that 
\begin{equation}\label{eq:surface-density}
\Sigma(r) =
\begin{cases}
\Sigma_0 \left({r}/r_\text{min}\right)^{-p} & r_{\text{min}} < r < r_{\text{max}},\\[2mm]
0 & \text{otherwise},
\end{cases}
\end{equation}
where $\Sigma_0$ is a reference surface density fixed by the total disk mass and the radii ($r_{\text{min}}$, $r_{\text{max}}$). 

\subsection{Initial conditions}

For each simulation, we uniformly select the total disk mass $M_{\text{disk}}/M_\star \in [0.5, 10] \times 10^{-4}$ and surface mass density power law exponent $p \in [0.5, 2)$ consistent with observations \citep{hartmann2008masses, williams2016measuring} and previous studies on intra-system uniformity \citep{weiss2022architectures}. We note that the heaviest disks in our simulation sample may be more massive than those inferred from observations, but such heavy disks are useful for understanding the physical processes at work. For simplicity, we choose the disk to orbit a star of mass $M_\star = 1\, M_\odot$.

Each simulation begins with a disk of planetary embryos on circular and coplanar orbits, consistent with eccentricity and inclination damping due to interaction with the protoplanetary disk \citep{kominami2002effect}. Each embryo has a mass uniformly chosen between $0.01\, M_\oplus$ and $0.1 \,M_\oplus$, comparable to values used in population synthesis and $N$-body studies \citep{morbidelli2015great,yzer2025forming}. The radius of each embryo is calculated by assuming a sphere of uniform density $\rho = 5\,\text{g}/\text{cm}^3$, though we note that our results are not sensitive to this choice (see Section~\ref{sec:assumptions} for $\rho = 2\,\text{g}/\text{cm}^3$).

We initialize each embryo on an orbit with semi-major axis chosen according to the probability distribution implied by the surface density power law $\Sigma \sim r^{-p}$. Specifically, for each embryo, a uniform random number $u$ is selected between $0$ and $1$, which is then used to calculate its initial semi-major axis via the inverse cumulative distribution function 
\begin{equation}
    r(u) = \left(r_{\text{min}}^{2-p} + u[r_{\text{max}}^{2-p} - r_{\text{min}}^{2-p}]\right)^{1/{(2-p)}},
\end{equation}
where we select $r_{\text{min}} = 0.05$ au and $r_{\text{max}} = 0.5$ au as the boundaries of the disk, in agreement with observed compact multi-planet systems \citep{winn2015occurrence}. The longitude of ascending node $\Omega$, argument of periapsis $\omega$, and mean anomaly $M$ of each orbit are each uniformly chosen between $0$ and $2\pi$.

The total number of planetary embryos in the simulation is determined by $N_e = \text{round}(M_{\text{disk}} / \overline{m}_{\text{embryo}})$, where $\overline{m}_{\text{embryo}} = 0.055\, M_\oplus$ is the average embryo mass. We note that our random sampling procedure generally produces total mass budgets which differ slightly from the prescribed disk masses. However, since $N_e \sim 10^3 -10^4$, this effect only amounts to a small ($\sim 1-2\%$) difference between the nominal and sampled disk masses.

\subsection{N-body simulations}

In order to compute statistics in various regions of parameter space, we carry out 1000 integrations of the aforementioned initial conditions, each for 1 Myr. As shown in Section~\ref{sec:results}, the vast majority of the mergers occur within the first $\sim 10$ kyr of the simulation, so that 1 Myr is generally sufficient for the system to reach a stable configuration.

Our simulations are carried out in the $N$-body code \texttt{REBOUND} \citep{rein2012rebound}. To efficiently handle collisions and close encounters, we use the hybrid integrator \texttt{TRACE} \citep{lu2024trace}, which uses the symplectic integrator \texttt{WHFast} \citep{rein2015whfast} when all orbiting bodies are separated by more than $4$ mutual Hill radii, and switches to the precise adaptive timestep integrator \texttt{IAS15} otherwise \citep{rein2015ias15}. 

During the integration, collisions are treated as perfect inelastic mergers. Specifically, when two bodies collide (i.e., when their separation is less than the sum of their radii), they are replaced by a single body whose mass is the sum of the two constituent masses. The velocity of the resulting body is calculated such that linear and angular momentum are conserved. The radius of the post-collision body is re-calculated assuming a constant density of $\rho = 5\,\text{g}/\text{cm}^3$. In this way, the orbital energy and gravitational self-energy of the resulting body is generally less than that of its constituents.

Bodies that become gravitationally unbound from their star are retained in the integration but excluded from the subsequent analysis. In addition, to quantify the number of instabilities which occur after the simulation window of $1$ Myr, we randomly select 50 of the 1000 previous integrations to run for an additional 9 Myr. As discussed below, longer-term effects do not substantially change our results. 

\section{Results}\label{sec:results}

\begin{figure}
    \centering
    \includegraphics[width=\linewidth]{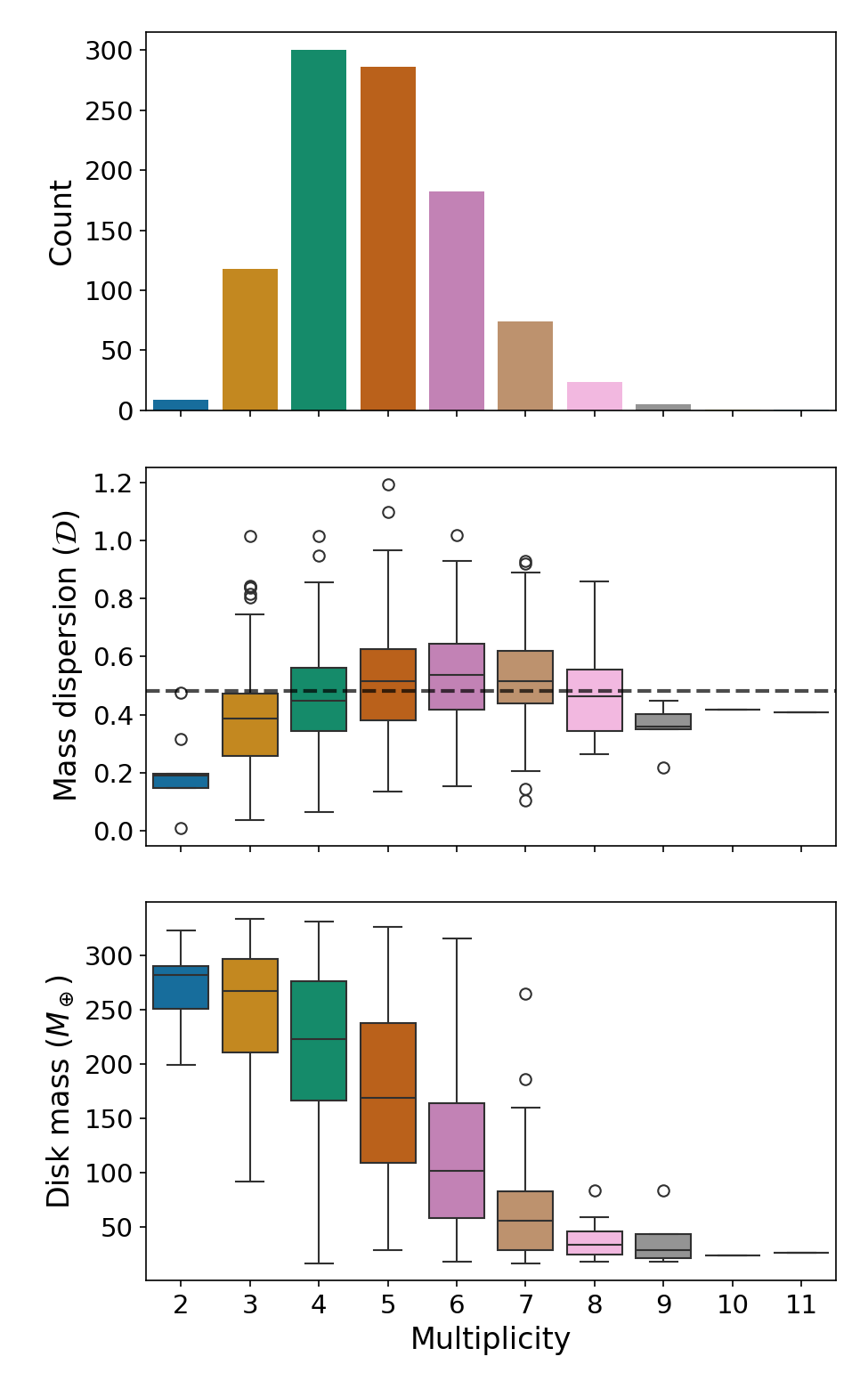}
    \caption{Mass dispersion $\mathcal{D}$ and disk mass as a function of post-instability multiplicity. Most systems have three to seven planets remaining after 1 Myr of evolution. Systems with heavier disks tend to form fewer planets. The dashed horizontal line in the middle panel indicates the horizontal line of best fit, $\mathcal{D} \approx 0.48$.}
    \label{fig:multiplicities}
\end{figure}

In the subsequent analysis, we neglect bodies which were ejected from each system during the integration. For the vast majority (97\%) of systems, these ejections amount to less than $\sim5\%$ of the original disk mass. Additionally, we neglect bodies that have final mass less than $1\, M_\oplus$.

\begin{figure}
    \centering
    \includegraphics[width=\linewidth]{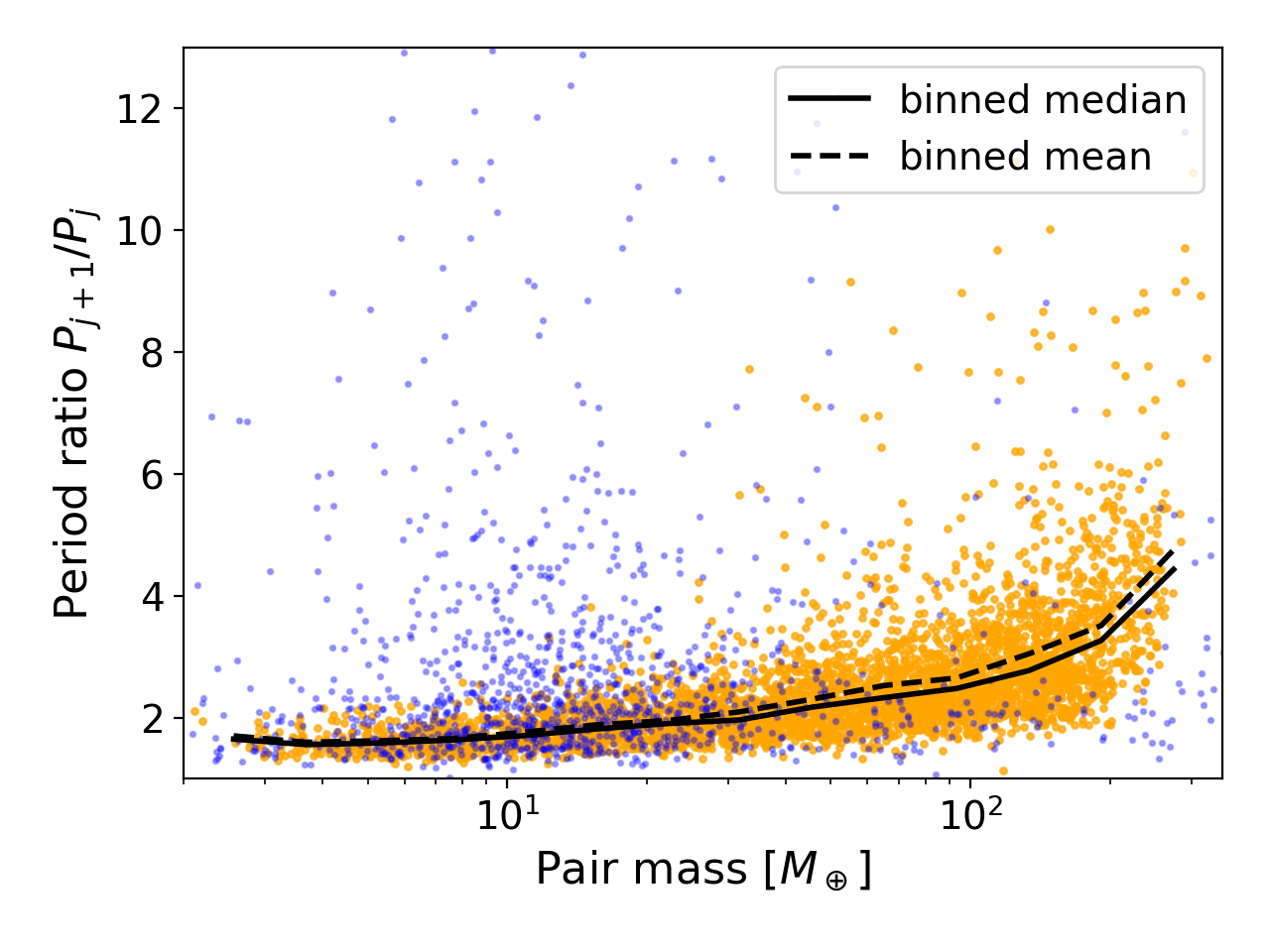}
    \caption{Period ratio vs. pair mass for planet pairs in our final simulated systems. The solid and dashed curves indicate the binned median and mean of the simulated data, respectively. The blue points indicate the period ratio vs. pair mass for planet pairs in observed systems (see Section~\ref{sec:observed}). The observed pairs often have somewhat higher period ratios at lower pair masses compared to the synthetic systems. }
    \label{fig:pratio}
\end{figure}

We find that the post-instability systems generally have multiplicity $N=3-7$ (see \Cref{fig:multiplicities}), consistent with observed compact multi-planet systems \citep{weiss2022architectures}. Systems with a high disk mass generally formed fewer ($N\sim 2-4$) high-mass planets, while systems with low disk mass generally formed more ($N\sim 4-7$) low-mass planets. The period ratio of planet pairs generally increases with the mass of the pair (\Cref{fig:pratio}), and we find no evidence of the formation of mean-motion resonances.

The planets in our sample range in mass from $\sim 1 M_\oplus$ to $\sim 100 M_\oplus$. We find that the masses of the inner and outer-most planet in each system are systematically smaller than the rest of the system. Specifically, the inner-most planet is the lightest in $\sim 44\%$ of systems, and likewise for the outermost planet. This mass ordering is substantially higher than what would be expected from a random ordering of planet masses ($1/N\sim 20\%$ for the median multiplicity $N=5$). We hypothesize that this trend arises from a narrower range of mass within the Hill sphere of the innermost and outermost protoplanets.

To evaluate the degree of intra-system uniformity, we use the mass dispersion metric $\mathcal{D} = \sigma_m / \overline{m}$, where $\overline{m}$ and $\sigma_m$ are the mean and standard deviation of the planet masses of the post-instability system, respectively. Smaller values of $\mathcal{D}$ correspond to systems which are more uniform, and $\mathcal{D}$ generally ranges from $\sim0.2 - 0.7$ for observed systems \citep{goldberg2022architectures}. We may analogously define the period ratio dispersion metric $\mathcal{D}_p = \sigma_\phi / \overline{\phi}$, where $\phi_j = P_{j+1}/P_j$ denotes the period ratio between successive planets, and $\overline{\phi}$ and $\sigma_\phi$ are the mean and standard deviation of the set of period ratios, respectively.

\begin{figure}
    \centering
    \includegraphics[width=\linewidth]{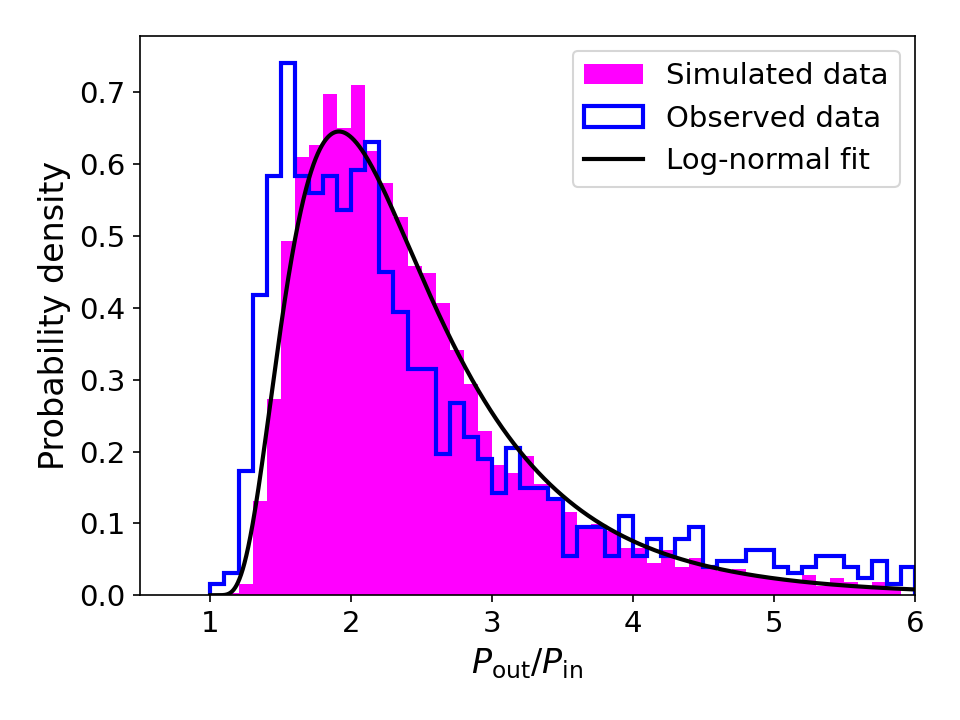}
    \caption{Distribution of period ratios among neighboring planets in our simulated systems. The black line indicates a fit of the simulated data to a log-normal distribution. The blue histogram indicates the distribution of period ratios among observed planet pairs (see Section~\ref{sec:observed}). The period ratios of the synthetic and observed systems are generally consistent.}
    \label{fig:period-ratios}
\end{figure}

The distribution of the period ratios of neighboring planets in our sample is shown in \Cref{fig:period-ratios}. The distribution is peaked near $P_{\text{out}}/P_{\text{in}} \approx 2$ and is well described by a log-normal distribution. For $m_{\text{pair}}=m_{\text{in}}+m_{\text{out}} < 60\, M_\oplus$, the median period ratio is $P_{\text{out}}/P_{\text{in}} \approx 1.9$, whereas for $m_{\text{pair}}>60\, M_{\oplus}$, it is $\approx 2.7$. This finding is consistent with the expectation that pairs of heavier planets have greater Hill radii and thus form with higher degrees of separation.

\begin{figure}
    \centering
    \includegraphics[width=\linewidth]{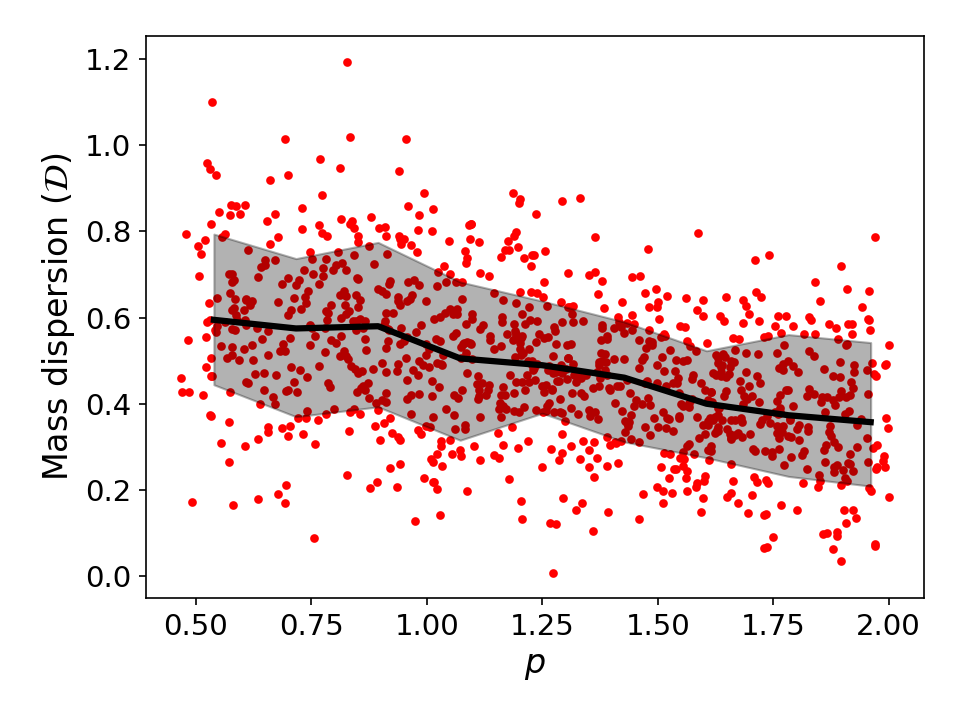}
    \caption{Mass dispersion $\mathcal{D}$ vs. surface density power law slope $p$ (\Cref{eq:surface-density}). The black line and shaded region indicate the binned median and the 16th-84th percentile range. Protoplanetary disks with steeper density profiles generally produce more uniform systems.}
    \label{fig:DvsP}
\end{figure}

We find that the mass dispersion of our post-instability systems is approximately $0.5$ for $N=3-7$ (\Cref{fig:multiplicities}), which is generally consistent with observed systems \citep{goldberg2022architectures}. Furthermore, we find a negative Pearson correlation ($R=-0.42$) between $\mathcal{D}$ and the surface mass density power law exponent $p$ (\Cref{fig:DvsP}). This trend is consistent with the finding that planetary systems with exactly equal mass and equal spacing have effective surface densities with $p=2$ \citep{adams2019pairwise}. We also note that this result is in agreement with \cite{he2022debiasing}, which found that $p=2$ is generally consistent with observed multi-planet systems. We find no strong correlation between $p$ and $\mathcal{D}_p$ ($R=-0.07$), nor between the disk mass and either $\mathcal{D}$ ($R=0.02$) or $\mathcal{D}_p$ ($R=0.25$). These null results suggest that mass uniformity is more important than period uniformity in the context of our study.

\begin{figure}
    \centering
    \includegraphics[width=\linewidth]{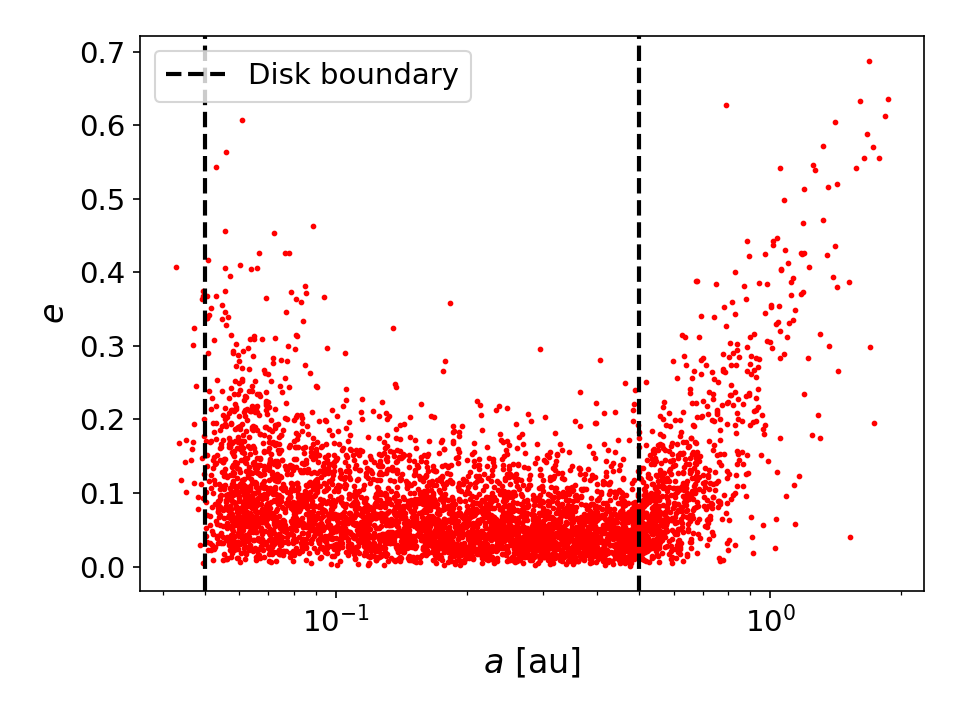}
    \caption{Eccentricity vs. semi-major axis for remaining planets. The vertical dashed lines indicate the disk boundaries $r_{\text{min}}=0.05\, \text{au}$ and $r_{\text{max}}=0.5\, \text{au}$.}
    \label{fig:aeplot}
\end{figure}

\Cref{fig:aeplot} shows the distribution of final eccentricities as a function of final semi-major axes. Planets that form within the initial boundaries of the disk $r_{\text{min}} < a < r_{\text{max}}$ are generally slightly eccentric $e \sim 0.1$. This finding is consistent with previous studies on the late stages of planet formation with \citep{duffell2015eccentric} and without \citep{chambers1998making} the presence of gas. Planets near or beyond the boundary ($a\gtrsim r_{\text{max}}$ or $a\lesssim r_{\text{min}}$) must arrive at those locations via scattering, and thus tend to exhibit higher eccentricities ($e\gtrsim 0.3$).

\subsection{Instability timescales}

\begin{figure}
    \centering
    \includegraphics[width=\linewidth]{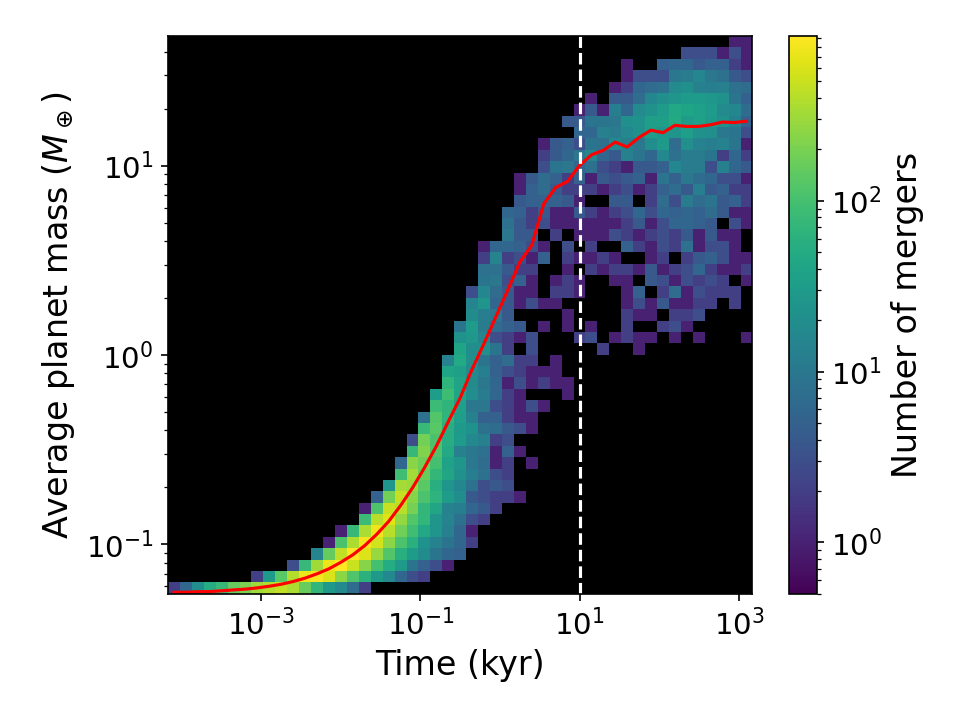}
    \caption{Heatmap showing the relative number of mergers over time. For each merger, the vertical axis indicates the average planet mass in the system immediately following the collision. The red line shows the average planet mass in each time bin. The vast majority of mergers occur within the first $\sim 10$ kyr of each simulation.}
    \label{fig:merger-map}
\end{figure}

Among the systems which were integrated for an additional 9 Myr, we find that most ($\sim 80^{+9}_{-13}\%$) remain stable for the full 10 Myr integration, with a small subset ($\sim 18^{+12}_{-8}\%$) undergoing a single additional merger, and even fewer ($\sim 2^{+8}_{-1}\%$) undergoing two additional mergers beyond those that occurred in the first Myr of evolution. The reported uncertainties are given by a 95\% Wilson score confidence interval \citep{wilson1927probable}.

For the complete set of 1000 simulations, \Cref{fig:merger-map} shows the total number of mergers as a function of time for the first 1 Myr of evolution for each simulation. The vast majority of mergers occur within the first $\sim 10$ kyr of the simulation as embryos assemble into more massive $(\gtrsim M_\oplus)$ planets. After 10 kyr, the number of mergers over time decreases substantially, indicating stability for the majority of systems.

\begin{figure}
    \centering
    \includegraphics[width=\linewidth]{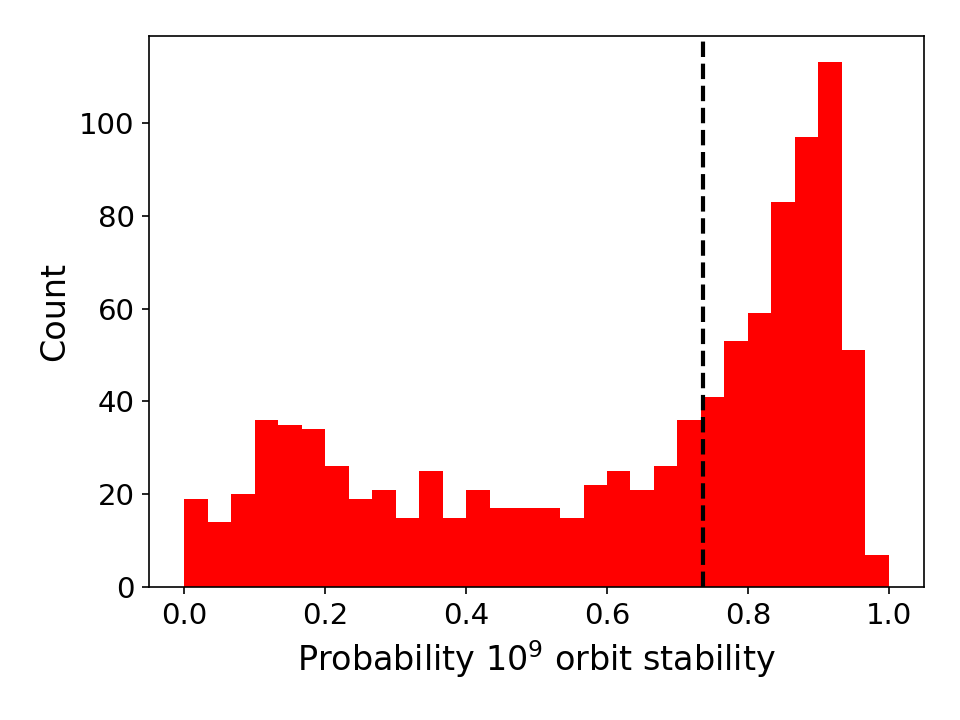}
    \caption{Distribution of stability probabilities for each simulated system, as inferred by \texttt{SPOCK}. Each probability indicates the likelihood that a given system remains stable for $10^9$ orbits of its innermost planet. The vertical black line indicates the median probability of stability.}
    \label{fig:spock-histogram}
\end{figure}

Additionally, we use the deep-learning tool \texttt{SPOCK} \citep{tamayo2020predicting, thadhani2025spock} to infer the long-term stability of the systems after 1 Myr of evolution. Specifically, we use the feature classifier to find that a majority of systems are inferred to have high likelihood to remain stable for more than $\sim 10^9$ orbits of their inner planet, corresponding to $10-500$ Myr (\Cref{fig:spock-histogram}).

Finally, we compute the mutual Hill separation between neighboring planets in the post-instability systems, defined by
\begin{equation}
    \Delta = \frac{a_2 - a_1}{R_H}
\end{equation}
where $a_1$ and $a_2$ are the semi-major axes of the inner and outer planets, and $R_H$, mutual Hill radius, is defined as
\begin{equation}
    R_H = \frac{a_1 + a_2}{2} \left(\frac{m_1 +m_2}{3M_\star}\right)^{1/3},
\end{equation}
where $m_1$ and $m_2$ are the mass of the inner and outer planets. We find that 91\% pairs have separation $\Delta \gtrsim 10$, indicating long-term stability \citep{gladman1993dynamics, smith2009orbital}.

\subsection{Bifurcation} 
\label{sec:bifurcation}

\begin{figure}
    \centering
    \includegraphics[width=\linewidth]{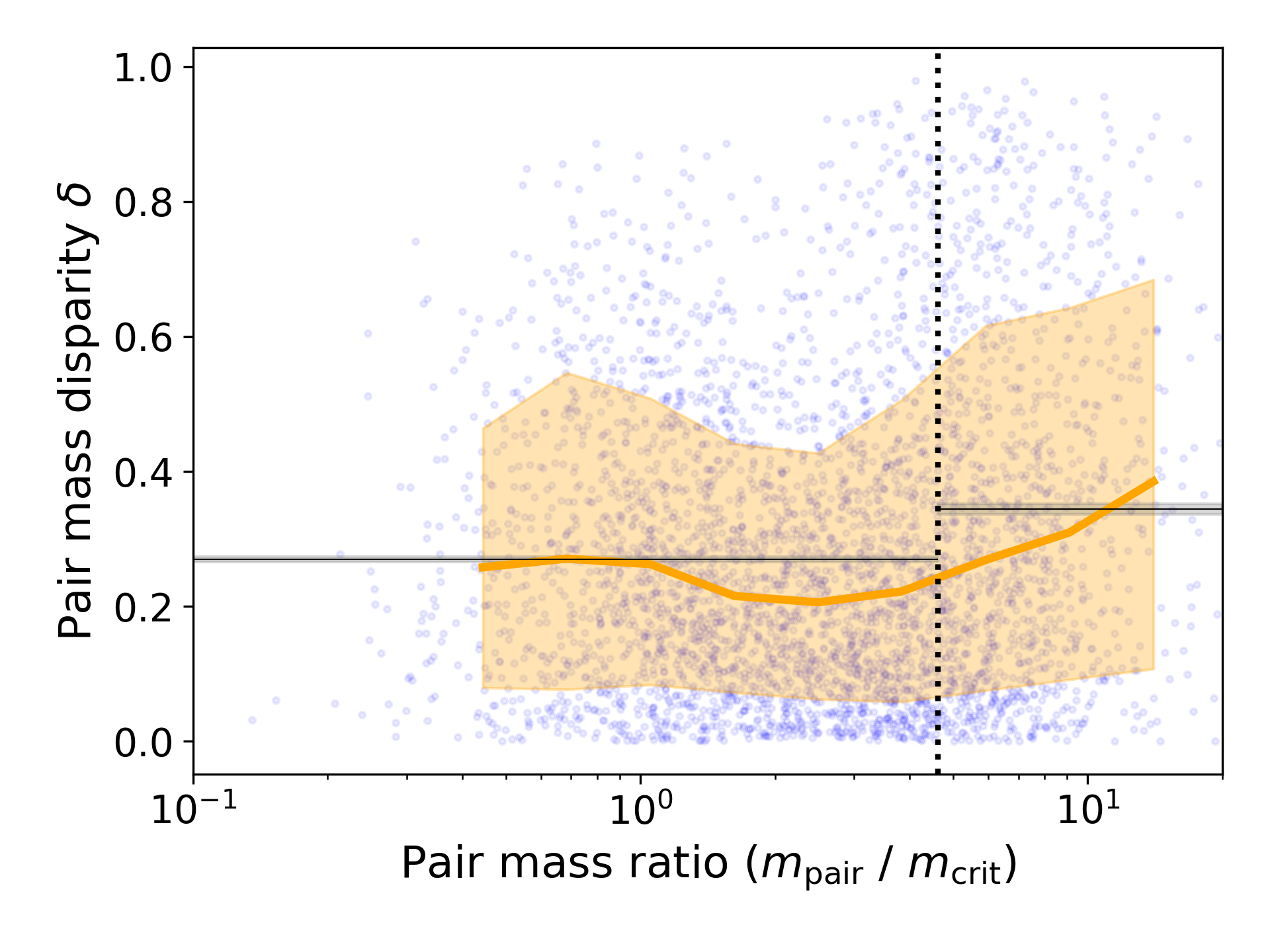}
    \caption{Pair mass disparity $(\delta)$ vs. ratio of pair mass to critical mass $m_{\text{crit}}$. The orange line and shaded region indicate the binned median and the 16th-84th percentile range. We choose $10$ bins of uniform size and bounds such that each bin has at least 100 planet pairs. For $m_1+m_2 > m_{\text{crit}}$, the mass disparity increases, as predicted by \citetalias{adams2020energy}. The vertical dotted line indicates the optimal cutoff determined by our $t$-test procedure, $m_1+m_2 = 4.6m_{\text{crit}}$. The horizontal lines and thin gray shaded regions surrounding them indicate the mean and standard error of the mass disparity for the low-mass and high-mass subsamples at the optimal cutoff.}
    \label{fig:mass-disparity}
\end{figure}

As shown in \citetalias{adams2020energy}, when a pair of planetary embryos exceeds a critical mass threshold $m_{\text{crit}}$, it becomes energetically favorable for one embryo to undergo runaway growth. In this section, we look for signs of this bifurcation from mass uniformity to runaway growth in our post-instability systems.

Specifically, runaway growth is energetically favored when the pair mass $m_\text{pair}$ satisfies (\citetalias{adams2020energy})
\begin{equation}
    m_{\text{pair}} > m_{\text{crit}} \approx M_\star \left(\frac{R_p}{a_2}\right) \frac{(\sqrt{\Lambda} + 2)(\sqrt{\Lambda}-1)^2}{4\alpha_g \sqrt{\Lambda}},
\end{equation}
where $R_p$ is the planet radius, $\alpha_g$ is an order-unity parameter related to the planet structure (see \citetalias{adams2020energy} for details), and $\Lambda = a_2/a_1$ where $a_1$ and $a_2$ are the semi-major axes of the inner and outer planets, respectively. We note that in \citetalias{adams2020energy}, $R_p$ and $\alpha_g$ are taken to be equal for both planets. We fix $\alpha_g = 0.5$ (the fiducial value used in \citetalias{adams2020energy}) in our analysis. Furthermore, in the subsequent analysis, we compute $R_p$ as the average radius between the two planets in the pair.

To differentiate between mass uniformity and runaway growth, we define the mass disparity between two neighboring planets to be 
\begin{equation}
    \delta = \frac{|m_1-m_2|}{m_1+m_2},
\end{equation}
where $m_1$ and $m_2$ are the masses of the inner and outer planets. Specifically, when the planet masses are approximately uniform ($m_1 \approx m_2$), we have $\delta \approx 0$, and when the masses are disparate ($m_1 \gg m_2$ or vice versa), $\delta \approx 1$.

\Cref{fig:mass-disparity} depicts the mass disparity $\delta$ as a function of the ratio between the pair mass $m_{\text{pair}} =m_1 + m_2$ and the threshold mass $m_{\text{crit}}$ for each pair of neighboring planets in the post-instability systems. We bin the data and compute the median mass disparity of each bin. We find that the mass disparity increases systematically as a function of $m_{\text{pair}}/m_{\text{crit}}$ once $m_{\text{pair}} > m_{\text{crit}}$, suggesting a decrease in mass uniformity for planet pairs above the critical mass threshold. However, we note that there are still many planet pairs with $m_1 \approx m_2$ but $m_1+m_2 > m_{\text{crit}}$, indicating that for some systems, the true minimum-energy configuration is not reached. We hypothesize that this mismatch is due to the fact that once embryos become sufficiently massive, mergers become rare and thus energy is not efficiently dissipated. We explore this further in Section~\ref{sec:dissipation}.

\begin{figure}
    \centering
    \includegraphics[width=\linewidth]{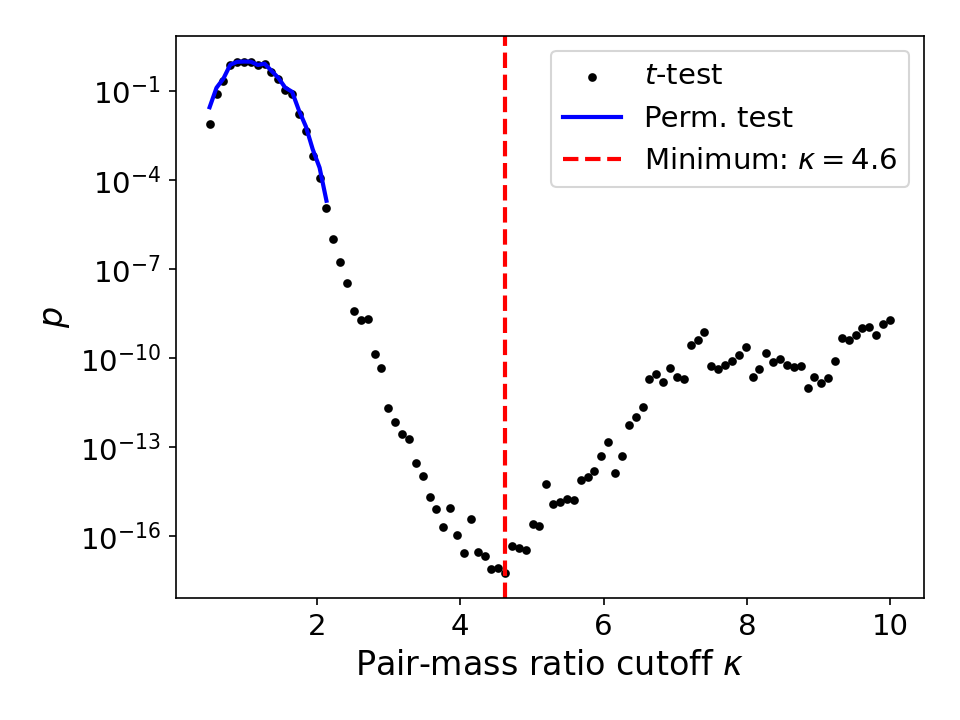}
    \caption{Results of the series of two-samples $t$-tests (black points) and permutation tests (blue line) on the mass disparity $\delta$, as a function of the pair-mass ratio cutoff $\kappa$. The minimum $p$-value is $3\times 10^{-18}$ when $\kappa =4.6$.}
    \label{fig:ttest}
\end{figure}

Additionally, we carry out a series of two-sample $t$-tests for various choices of the pair-mass ratio cutoff $\kappa = m_{\text{pair}}/m_{\text{crit}}$. Specifically, for various values of $\kappa \in [0.5,10]$, we split our sample of post-instability planets into two subsamples: those with $ m_{\text{pair}}/m_{\text{crit}} < \kappa$, and those with $m_{\text{pair}}/m_{\text{crit}} > \kappa$. For each choice of $\kappa$, we perform a two-sample $t$-test on the mass disparity $\delta$ between the two subsamples and plot the corresponding $p$-value (\Cref{fig:ttest}). We expect that for $\kappa\sim 1$, energy minimization implies that $\delta$ is distributed differently between the two samples. We find that this is true ($p\lesssim 0.01$) for $\kappa \gtrsim 2$, in agreement with the trend in \Cref{fig:mass-disparity}. We stress that the $p$-value at the optimal choice of $\kappa$ is $\sim 10^{-16}$ times smaller than that near $\kappa=1$. 

Furthermore, for each $t$-test, we carry out a non-parametric permutation test which makes minimal assumptions of the underlying distribution \citep{edgington2007randomization}. Specifically, for each $\kappa$, we randomly permute the mass disparities of each pair of subsamples, assigning a ``low-mass" or ``high-mass" label to each element. We carry out $N_p=10^5$ permutations. The $p$-value is calculated as $(b+1)/(N_p+1)$ where $b$ is the number of permutations whose difference in mean mass disparity exceeds that of the fiducial subsamples for a given $\kappa$. The results of the permutation tests are shown in \Cref{fig:ttest} and match closely to the results of the $t$-tests. Since the permutation test can only attain a minimum $p$-value of $1/(N_p+1)\approx 10^{-5}$, the tests are only reliable for $\kappa \lesssim 2$.

For the optimal cutoff $\kappa = 4.6$, the mean mass disparity of the low-mass sample is $0.269 \pm 0.004$ and the mean mass disparity for the high-mass sample is $0.335 \pm 0.007$. Although the mean mass disparities only differ by $\sim25\%$, this difference represents $\sim 8$ standard errors, indicating a significant trend.

This finding indicates that planetary systems start to depart from mass uniformity when their total mass exceeds the threshold, but full runaway behavior does not occur until the pair mass is a factor of $\sim 4$ higher than the critical threshold of \citetalias{adams2020energy}. On the other hand, this discrepancy may suggest that the fiducial parameters of \citetalias{adams2020energy} are off by a factor of a few, and/or indicate the necessity to include additional astrophysical channels for energy dissipation.

\subsection{Efficiency of energy dissipation}\label{sec:dissipation}

\begin{figure*}
    \centering
    \includegraphics[width=\linewidth]{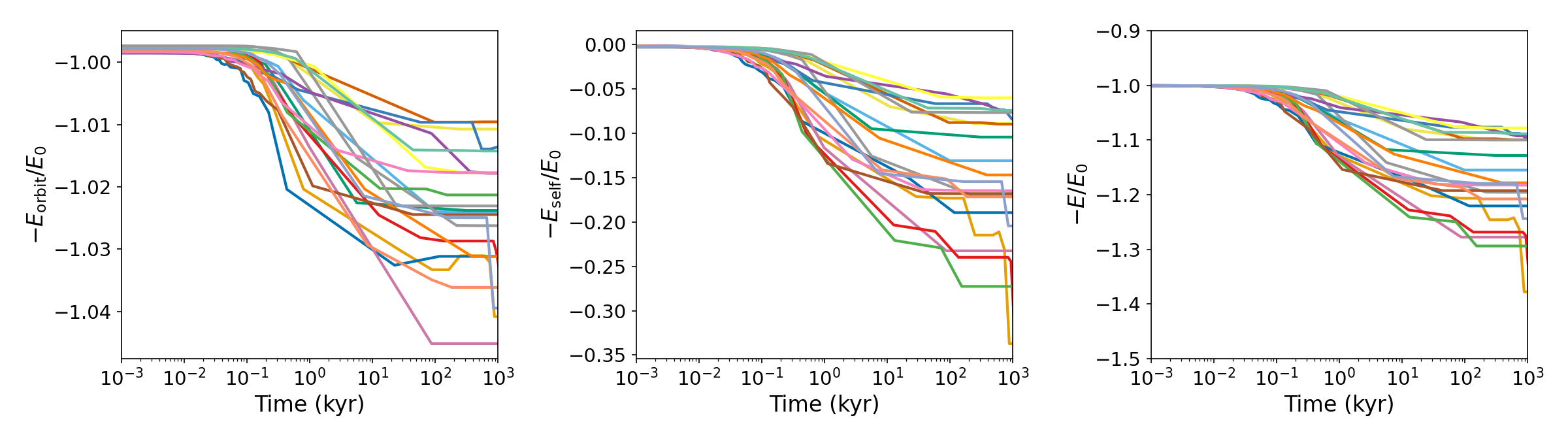}
    \caption{Orbital energy $E_{\text{orbit}}=-\sum_{i} GM_\star m_i/2a_i$ (left), gravitational self-energy $E_{\text{self}}=-\sum_i(3/5)Gm_i^2/R_i$ (middle), and total energy $E=E_{\rm orbit} + E_{\rm self}$(right) over time for several simulated systems, each normalized by the initial simulation energy $E_0$. Here, the sums are over all bound bodies, and $m_i$ and $R_i$ are the mass and radius of the $i^{\rm th}$ body. The majority of energy dissipation occurs over the first $\sim 10$ kyr, consistent with \Cref{fig:merger-map}. Systems generally lose $10-30\%$ of their initial energy before attaining an approximately stable configuration.}
    \label{fig:energy-dynamics}
\end{figure*}

In order for a system of planetary embryos to reach the minimum energy configuration, the energy dissipation from inelastic collisions must be rapid enough and persist for long enough for the system to dissipate large fractions of its total energy. 

In \Cref{fig:energy-dynamics}, we plot the orbital energy, gravitational self-energy, and total energy over time for several simulated systems. As suggested by \Cref{fig:merger-map}, the vast majority of mergers occur within the first $\sim 10$ kyr, and thus the majority of energy dissipation occurs on this timescale (\Cref{fig:energy-dynamics}). After $\sim 10\, \text{kyr}$, the system evolves to an approximately Hill-stable configuration, mergers become rare, and energy dissipation effectively shuts off.

For each system, we compute the characteristic timescale for energy dissipation $\tau = |E/ \dot{E}|$; we find that during the period with the most rapid energy dissipation ($t < 10$ kyr), $\tau$ is generally $\sim 10$ kyr. Thus, systems generally dissipate $\sim10-30\%$ of their energy, but do not always have time to efficiently dissipate enough energy to reach the minimum energy configuration.

Notice also that the dissipation of orbital energy (left panel of Figure \ref{fig:energy-dynamics}) corresponds to a change of only 1 -- 4\%. The corresponding energy change required to reach the energy-optimized state is somewhat larger, of order 10\% (see Figure 1 of \citealt{adams2019pairwise}). This discrepancy, again, indicates that the simulated systems do not always have time to fully reach their minimum energy configurations. 

\section{Discussion}\label{sec:discussion}
We have shown that the rapid accumulation of rocky planetary embryos generally yields mass uniformity for $m_{\text{pair}} < m_{\text{crit}}$ and mass disparity for $m_{\text{pair}} > m_{\text{crit}}$. In this section, we discuss the limitations of our model and highlight the applications of the energy minimization hypothesis to the observed exoplanet sample.

\subsection{Model assumptions}\label{sec:assumptions}
First, we highlight that our assumption of a uniform density of $\rho = 5\, \text{g}/\text{cm}^3$ is not expected to hold across the range of planet masses in our sample $\sim1-100\, M_\oplus$. Specifically, we expect that as a planet's mass increases, it eventually begins to accrete gas, resulting in lower density. This change of density is manifested in different mass-radius relations for various mass regimes \citep{chen2017probabilistic}. For planets with lower densities (and thus greater radii for the same mass), we expect the effective cross section of the collision to be greater than that assumed in our model, resulting in a higher rate of collisions than found in our simulations.

To evaluate the robustness of our model, we carry out an additional 200 integrations which adopt a constant density of $\rho = 2\, \text{g}/\text{cm}^3$, closer to what is expected of gas giants. We find that the reported trends in mass disparity (\Cref{fig:mass-disparity}) and mass dispersion (\Cref{fig:DvsP}, \Cref{fig:mass-disparity}) remain in the sample with lower-density planetary embryos. Systems in this sample typically had final multiplicity $N=3-7$. The best-fit mass dispersion $\mathcal{D}\approx 0.45$ for this sample is slightly lower than that of the fiducial sample. As in the $\rho = 5\, \text{g}/\text{cm}^3$ sample, the typical eccentricity for planets in the disk region $r_{\text{min}} < a < r_{\text{max}}$ was $e\sim 0.1$, and planets outside of the disk region had $e \gtrsim 0.3$. Likewise, the distribution of $P_{\text{out}}/P_{\text{in}}$ is well-approximated by a log-normal distribution peaked at $\sim 2$. We note that the critical pair-mass ratio cutoff inferred from the $t$-test procedure performed in Section~\ref{sec:bifurcation} is slightly higher in the low-density sample ($\kappa = 5.5$). This corresponds to a difference in minimal $p$-value of about five orders of magnitude compared to the fiducial sample, due to the smaller sample size.

\begin{figure}
    \centering
    \includegraphics[width=\linewidth]{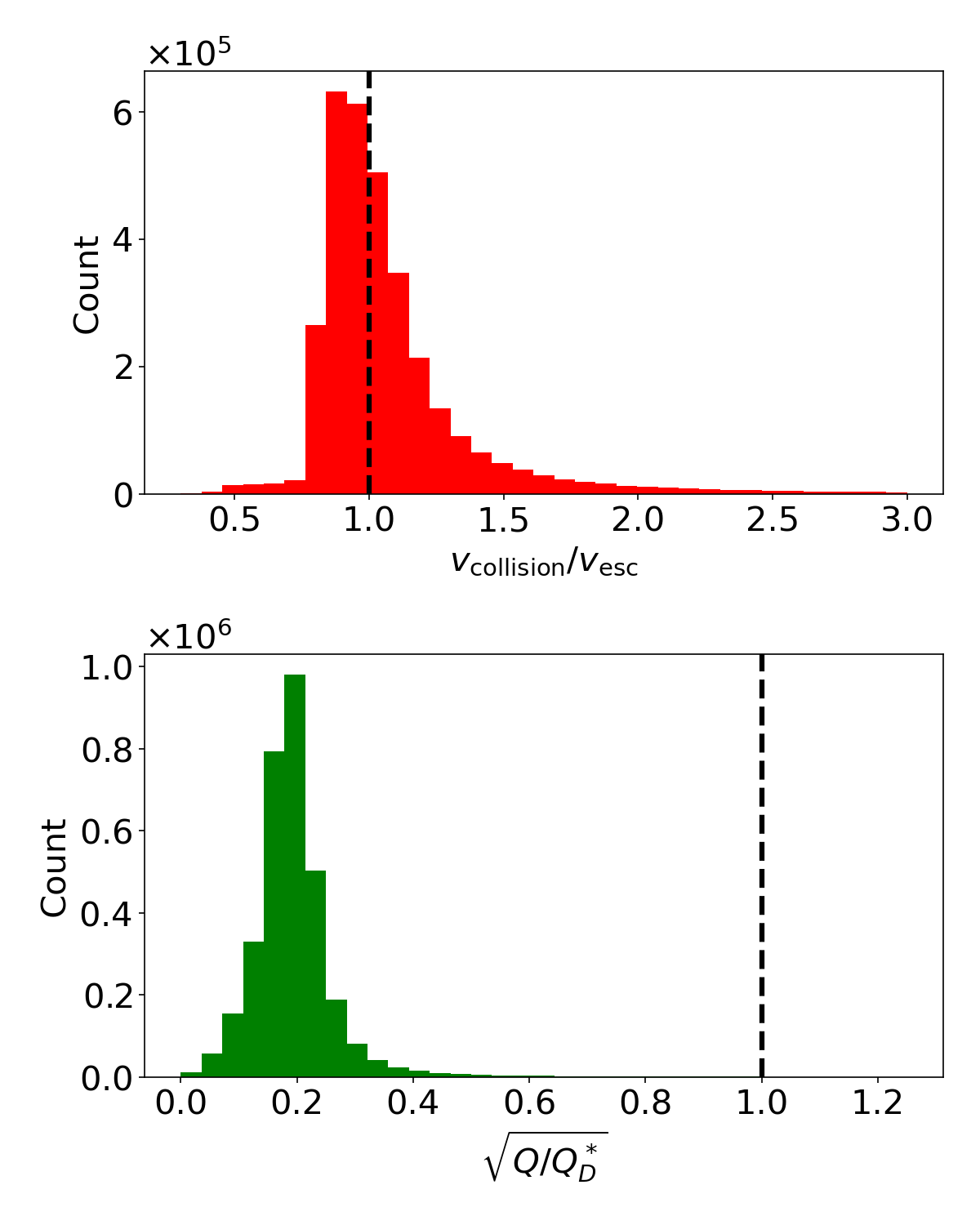}
    \caption{Top panel: distribution of $v_{\text{collision}}/v_{\text{esc}}$ for all collisions in the simulation sample; the black vertical line indicates $v_{\text{collision}} = v_{\text{esc}}$. Bottom panel: distribution of $\sqrt{Q/Q_D^*}$ for all collisions in the simulation sample; the black vertical line indicates $Q = Q_D^{*}$. We find that the collisions are not generally violent enough to induce fragmentation.}
    \label{fig:collision-velocities}
\end{figure}

Furthermore, our model assumes that collisions are perfect inelastic mergers, which may not be true in real systems. To assess the validity of this assumption, we proceed by analyzing the distribution of impact velocities following the procedure of \cite{goldberg2022architectures}. Specifically, for a given collision of two objects of masses $m$ and $m'$ and radius $R$ and $R'$ and impact angle $\theta$, only a fraction of the masses of the two bodies interact. The interaction mass is roughly approximated by \citep{leinhardt2012collisions}
\begin{equation}
    m'_{\text{int}} \approx \frac{3R' \ell^2 - \ell^3}{4R'^3} m',
\end{equation}
where the projected length $\ell$ is
\begin{equation}
    \ell = (R+R')(1-\sin{\theta}).
\end{equation}
Then, perfect mergers are expected whenever the velocity of the collision, $v_{\text{collision}}$, is less than the escape velocity of the post-collision body \citep{stewart2012collisions}:
\begin{equation}
    v_{\text{esc}} = \sqrt{2G(m+m_\text{int}')/R_\text{new}},
\end{equation}
where $G$ is the gravitational constant and $R_{\text{new}}$ is the radius of the newly-formed object (in our case, calculated by assuming a fixed density $\rho = 5 \,\text{g}/\text{cm}^3$).

Additionally, colliding bodies are expected to fragment when the specific energy of the collision
\begin{equation}
    Q = \frac{m' v_{\text{collision}}^2}{2(m+m')}
\end{equation}
is sufficiently high to unbind the mass of one body into two or more objects. Empirically, in the gravity-dominated regime, this occurs when $Q$ exceeds the critical disruption threshold
\begin{equation}
    Q_D^* = q_g \rho \left(\frac{R}{1\,\text{cm}}\right)^b,
\end{equation}
with $q_g \approx 0.5 \text{ erg}\, \text{cm}^3\, \text{g}^{-2}$ and $b = 1.36$ \citep{armitage2020astrophysics}.

We plot the distributions of $v_{\text{collision}}/v_{\text{esc}}$ and $\sqrt{Q/Q_D^*}$ for all collisions in each simulation in our sample in \Cref{fig:collision-velocities}. We find that most collisions have $v_{\rm collision} \sim v_{\rm esc}$ within a factor of $\sim 2$, indicating that many collisions are imperfect mergers and thus may eject material \citep{goldberg2022architectures}. In contrast, we find that $Q < Q_D^*$ for $99\%$ of collisions, and therefore we expect fragmentation to be negligible in our simulations. We thus find that the assumption of perfect inelastic mergers is reasonable. Collisions with $Q > Q_D^*$ or $v_{\text{collision}} \gg v_{\text{esc}}$ generally occur when the impact angle is high ($\theta \gtrsim 90 \deg$), since in this case, the relative velocity between the colliding bodies is large.

\subsection{Application to observed systems}\label{sec:observed}

\begin{figure}
    \centering
    \includegraphics[width=\linewidth]{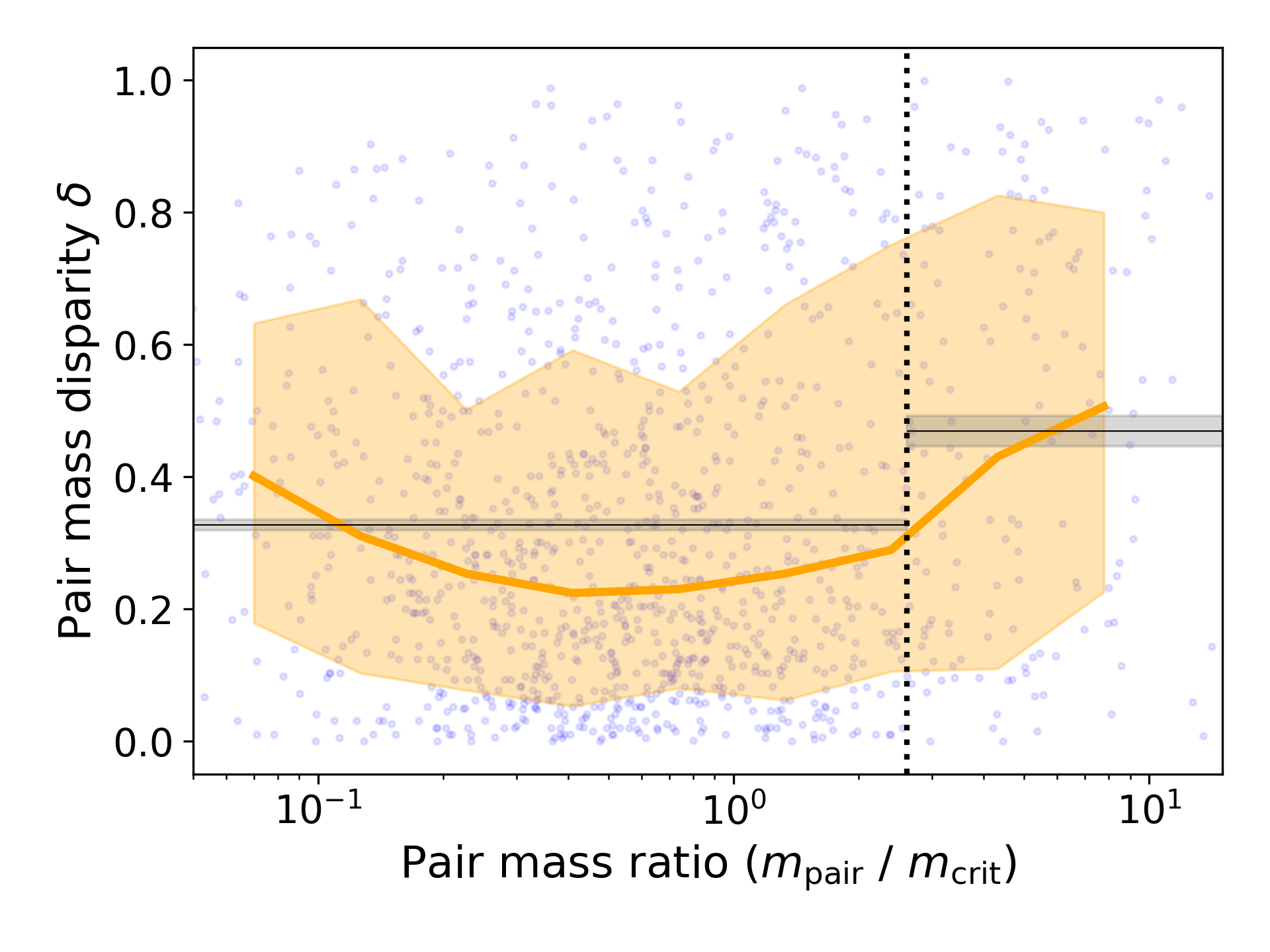}
    \caption{Same as \Cref{fig:mass-disparity}, but for observed exoplanet systems. Again, for $m_1 + m_2 > m_{\text{crit}}$, the mass disparity increases. The vertical dotted line indicates the optimal cutoff determined by our $t$-test procedure, $m_1+m_2 = 2.6m_{\text{crit}}$. The horizontal lines and thin gray shaded regions surrounding them indicate the mean and standard error of the mass disparity for the low-mass and high-mass subsamples at the optimal cutoff.}
    \label{fig:bifurcation-observed}
\end{figure}

To understand the energy minimization hypothesis in the context of observed systems, we plot the mass disparity $\delta$ vs. the pair mass $m_\text{pair}$ for observed systems from the NASA Exoplanet Archive \citep{akeson2013astro} in \Cref{fig:bifurcation-observed}. For planets without a radius (mass) measurement, we infer the radius (mass) from the probabilistic mass-radius relation \texttt{Forecaster} \citep{chen2017probabilistic}. We find that the trend from \Cref{fig:mass-disparity} is also present among observed systems, with the mass disparity $\delta$ growing as a function of $m_{\text{pair}}$ for $m_{\text{pair}} > {m_\text{crit}}$. 

A two-sample $t$-test of the distribution of mass disparities between planet pairs with $m_{\text{pair}} > m_{\text{crit}}$ and $m_{\text{pair}} < m_{\text{crit}}$ yields a $p$-value of $5.6\times 10^{-6}$. Thus, we conclude that observed planet pairs with $m_{\text{pair}} > m_{\text{crit}}$ are generally less uniform than those with  $m_{\text{pair}} < m_{\text{crit}}$.

Through the same ensemble of $t$-tests carried out for the synthetic data, we find that the optimal pair-mass ratio for observed planet pairs is $\kappa = 2.6$. Additionally, we repeat this analysis using the mass-radius relationship provided by \cite{muller2024mass} and find an optimal pair-mass ratio $\kappa=2.2$. Furthermore, we carry out a suite of permutation tests analogous to those described in Sec~\ref{sec:bifurcation} and find overall agreement. These values are markedly lower than the value obtained from the analysis of the synthetic systems, and may be indicative of additional astrophysical mechanisms responsible for energy dissipation among planet pairs, such as migration and eccentricity damping due to interactions with the protoplanetary disk, and pebble accretion.

At the optimal cutoff $\kappa=2.6$, the mean mass disparity of the low-mass sample is $0.341\pm 0.013$ and the mean mass disparity of the high-mass sample is $0.458 \pm 0.024$. This amounts to a difference of $\sim 4$ standard errors, which is still statistically robust, but the signal to noise ratio is smaller than for the synthetic data.

\section{Conclusion}\label{sec:conclusion}

The transition from intra-system mass uniformity to non-uniformity for systems with multiple planets is a key problem in understanding the formation and dynamics of planetary systems. In this paper, we carried out a suite of $N$-body integrations to evaluate the energy minimization hypothesis for planet formation (from \citetalias{adams2020energy}), namely, that the observed trends in uniformity are a result of nearly uniform configurations being the lowest energy state accessible to forming planetary pairs. Our numerical results show that planet pairs above the critical mass threshold derived in \citetalias{adams2020energy} tend to favor less uniform configurations and we confirmed that this trend is present in the sample of observed exoplanets. We also found that in our simulations, energy dissipation occurs rapidly, but only while mergers are common. After $\sim 10$ kyr, systems become approximately Hill stable and mergers become rare. As a result, the global energy minimum state is not always reached. This finding indicates a more complicated picture: planet pairs below the threshold mass are energetically favored to have nearly equal masses and generally reach that state. When the pair mass is just above threshold, the pairs start to favor unequal masses, but the full realization of runaway growth (where one planet dominates the mass budget) does not occur until a mass scale $\sim4$ times above the threshold mass of \citetalias{adams2020energy}. For comparison, the critical mass threshold favored by the population of observed planet pairs is $\sim 2-3$ larger than that of the analytical threshold. 

While this paper provides numerical evidence for energy optimization's role in planet formation, a great deal of additional work should be carried out. One compelling direction for future analytical work is to modify the argument of \citetalias{adams2020energy} to account for incomplete energy minimization. Future $N$-body studies could improve our model by accounting for fragmentation in high-energy collisions, self-consistently modeling the radius evolution of the planetary embryos and planets over time, and additional channels for energy dissipation, such as interactions with the protoplanetary disk and pebble accretion. In addition, it would be interesting to repeat our numerical simulations with lighter embryos $\lesssim 0.01\, M_\oplus$ to determine how well the reported trends persist.

\bigskip
\centerline{Acknowledgments}
\medskip 

We thank the anonymous reviewer for comments that have substantially improved our manuscript. We thank Fei Dai for helpful suggestions. This material is based on work supported by the National Science Foundation Graduate Research Fellowship Program. The authors acknowledge the MIT Office of Research Computing and Data and the MIT Engaging Cluster for providing computational resources that contributed to the results reported in this paper. FCA is also supported by NSF Grant No. 2508843, and by the Leinweber Institute for Theoretical Physics at the University of Michigan. This research made use of the NASA Exoplanet Archive, which is operated by the California Institute of Technology, under contract with the National Aeronautics and Space Administration under the Exoplanet Exploration Program.

\bibliography{biblio}{}

@article{rein2012rebound,
  title={REBOUND: an open-source multi-purpose N-body code for collisional dynamics},
  author={Rein, Hanno and Liu, S-F},
  journal={Astronomy \& Astrophysics},
  volume={537},
  pages={A128},
  year={2012},
  publisher={EDP Sciences}
}

@article{lu2024trace,
  title={TRACE: a code for time-reversible astrophysical close encounters},
  author={Lu, Tiger and Hernandez, David M and Rein, Hanno},
  journal={Monthly Notices of the Royal Astronomical Society},
  volume={533},
  number={3},
  pages={3708--3723},
  year={2024},
  publisher={Oxford University Press}
}

@article{rein2015whfast,
  title={WHFAST: a fast and unbiased implementation of a symplectic Wisdom--Holman integrator for long-term gravitational simulations},
  author={Rein, Hanno and Tamayo, Daniel},
  journal={Monthly Notices of the Royal Astronomical Society},
  volume={452},
  number={1},
  pages={376--388},
  year={2015},
  publisher={Oxford University Press}
}

@article{rein2015ias15,
  title={IAS15: a fast, adaptive, high-order integrator for gravitational dynamics, accurate to machine precision over a billion orbits},
  author={Rein, Hanno and Spiegel, David S},
  journal={Monthly Notices of the Royal Astronomical Society},
  volume={446},
  number={2},
  pages={1424--1437},
  year={2015},
  publisher={Oxford University Press}
}

@article{weiss2022architectures,
  title={Architectures of Compact Multi-Planet Systems: Diversity and Uniformity},
  author={Weiss, LM and Millholland, SC and Petigura, EA and Adams, FC and Batygin, K and Block, AM and Mordasini, C},
  journal={Protostars and Planets VII},
  volume={534},
  pages={863},
  year={2023}
}

@article{williams2016measuring,
  title={Measuring protoplanetary disk gas surface density profiles with ALMA},
  author={Williams, Jonathan P and McPartland, Conor},
  journal={The Astrophysical Journal},
  volume={830},
  number={1},
  pages={32},
  year={2016},
  publisher={The American Astronomical Society}
}

@article{hartmann2008masses,
  title={Masses and mass distributions of protoplanetary disks},
  author={Hartmann, Lee},
  journal={Physica Scripta},
  volume={2008},
  number={T130},
  pages={014012},
  year={2008}
}

@article{morbidelli2015great,
  title={The great dichotomy of the Solar System: Small terrestrial embryos and massive giant planet cores},
  author={Morbidelli, A and Lambrechts, Michiel and Jacobson, S and Bitsch, Bertram},
  journal={Icarus},
  volume={258},
  pages={418--429},
  year={2015},
  publisher={Elsevier}
}

@article{yzer2025forming,
  title={Forming Earth-like and low-mass rocky exoplanets through pebble and planetesimal accretion},
  author={Yzer, Mitchell and Brasser, Ramon and Ten Kate, Inge Loes},
  journal={Astronomy \& Astrophysics},
  volume={698},
  pages={A307},
  year={2025},
  publisher={EDP Sciences}
}

@article{kominami2002effect,
  title={The effect of tidal interaction with a gas disk on formation of terrestrial planets},
  author={Kominami, Junko and Ida, Shigeru},
  journal={Icarus},
  volume={157},
  number={1},
  pages={43--56},
  year={2002},
  publisher={Elsevier}
}

@article{tamayo2020predicting,
  title={Predicting the long-term stability of compact multiplanet systems},
  author={Tamayo, Daniel and Cranmer, Miles and Hadden, Samuel and Rein, Hanno and Battaglia, Peter and Obertas, Alysa and Armitage, Philip J and Ho, Shirley and Spergel, David N and Gilbertson, Christian and others},
  journal={Proceedings of the National Academy of Sciences},
  volume={117},
  number={31},
  pages={18194--18205},
  year={2020},
  publisher={National Academy of Sciences}
}

@article{thadhani2025spock,
  title={SPOCK 2.0: Updates to the FeatureClassifier in the Stability of Planetary Orbital Configurations Klassifier},
  author={Thadhani, Elio and Ba, Yanming and Rein, Hanno and Tamayo, Daniel},
  journal={Research Notes of the AAS},
  volume={9},
  number={2},
  pages={27},
  year={2025},
  publisher={The American Astronomical Society}
}

@article{gladman1993dynamics,
  title={Dynamics of systems of two close planets},
  author={Gladman, Brett},
  journal={Icarus},
  volume={106},
  number={1},
  pages={247--263},
  year={1993},
  publisher={Elsevier}
}

@article{adams2020energy,
  title={Energy optimization in extrasolar planetary systems: the transition from peas-in-a-pod to runaway growth},
  author={Adams, Fred C and Batygin, Konstantin and Bloch, Anthony M and Laughlin, Gregory},
  journal={Monthly Notices of the Royal Astronomical Society},
  volume={493},
  number={4},
  pages={5520--5531},
  year={2020},
  publisher={Oxford University Press}
}

@article{goldberg2022architectures,
  title={Architectures of compact super-Earth systems shaped by instabilities},
  author={Goldberg, Max and Batygin, Konstantin},
  journal={The Astronomical Journal},
  volume={163},
  number={5},
  pages={201},
  year={2022},
  publisher={The American Astronomical Society}
}

@article{chen2017probabilistic,
  title={Probabilistic forecasting of the masses and radii of other worlds},
  author={Chen, Jingjing and Kipping, David},
  journal={The Astrophysical Journal},
  volume={834},
  number={1},
  pages={17},
  year={2017},
  publisher={The American Astronomical Society}
}

@article{leinhardt2012collisions,
  title={Collisions between gravity-dominated bodies. I. Outcome regimes and scaling laws},
  author={Leinhardt, Zo{\"e} M and Stewart, Sarah T},
  journal={The Astrophysical Journal},
  volume={745},
  number={1},
  pages={79},
  year={2012},
  publisher={The American Astronomical Society}
}

@article{stewart2012collisions,
  title={Collisions between gravity-dominated bodies. II. The diversity of impact outcomes during the end stage of planet formation},
  author={Stewart, Sarah T and Leinhardt, Zo{\"e} M},
  journal={The Astrophysical Journal},
  volume={751},
  number={1},
  pages={32},
  year={2012},
  publisher={The American Astronomical Society}
}

@book{armitage2020astrophysics,
  title={Astrophysics of planet formation},
  author={Armitage, Philip J},
  year={2020},
  publisher={Cambridge University Press}
}

@article{millholland2017kepler,
  title={Kepler multi-planet systems exhibit unexpected intra-system uniformity in mass and radius},
  author={Millholland, Sarah and Wang, Songhu and Laughlin, Gregory},
  journal={The Astrophysical Journal Letters},
  volume={849},
  number={2},
  pages={L33},
  year={2017},
  publisher={The American Astronomical Society}
}

@article{weiss2018california,
  title={The California-Kepler Survey. V. Peas in a pod: planets in a Kepler multi-planet system are similar in size and regularly spaced},
  author={Weiss, Lauren M and Marcy, Geoffrey W and Petigura, Erik A and Fulton, Benjamin J and Howard, Andrew W and Winn, Joshua N and Isaacson, Howard T and Morton, Timothy D and Hirsch, Lea A and Sinukoff, Evan J and others},
  journal={The Astronomical Journal},
  volume={155},
  number={1},
  pages={48},
  year={2018},
  publisher={The American Astronomical Society}
}

@article{weiss2018california2,
  title={The California-Kepler Survey. VI. Kepler Multis and Singles Have Similar Planet and Stellar Properties Indicating a Common Origin},
  author={Weiss, Lauren M and Isaacson, Howard T and Marcy, Geoffrey W and Howard, Andrew W and Petigura, Erik A and Fulton, Benjamin J and Winn, Joshua N and Hirsch, Lea and Sinukoff, Evan and Rowe, Jason F and others},
  journal={The Astronomical Journal},
  volume={156},
  number={6},
  pages={254},
  year={2018},
  publisher={The American Astronomical Society}
}

@article{wang2017rv,
  title={RV-detected Kepler-multi Analogs Exhibit Intra-system Mass Uniformity},
  author={Wang, Songhu},
  journal={Research Notes of the American Astronomical Society},
  volume={1},
  number={1},
  pages={26},
  year={2017}
}

@article{tremaine2015statistical,
  title={The statistical mechanics of planet orbits},
  author={Tremaine, Scott},
  journal={The Astrophysical Journal},
  volume={807},
  number={2},
  pages={157},
  year={2015},
  publisher={The American Astronomical Society}
}

@article{adams2019pairwise,
  title={Pairwise tidal equilibrium states and the architecture of extrasolar planetary systems},
  author={Adams, Fred C},
  journal={Monthly Notices of the Royal Astronomical Society},
  volume={488},
  number={1},
  pages={1446--1461},
  year={2019},
  publisher={Oxford University Press}
}

@article{mishra2021new,
  title={The New Generation Planetary Population Synthesis (NGPPS) VI. Introducing KOBE: Kepler Observes Bern Exoplanets-Theoretical perspectives on the architecture of planetary systems: Peas in a pod},
  author={Mishra, Lokesh and Alibert, Yann and Leleu, Adrien and Emsenhuber, Alexandre and Mordasini, Christoph and Burn, Remo and Udry, St{\'e}phane and Benz, Willy},
  journal={Astronomy \& Astrophysics},
  volume={656},
  pages={A74},
  year={2021},
  publisher={EDP Sciences}
}

@article{lammers2023intra,
  title={Intra-system uniformity: a natural outcome of dynamical sculpting},
  author={Lammers, Caleb and Hadden, Sam and Murray, Norman},
  journal={Monthly Notices of the Royal Astronomical Society: Letters},
  volume={525},
  number={1},
  pages={L66--L71},
  year={2023},
  publisher={Oxford University Press}
}

@ARTICLE{akeson2013astro,
       author = {{Akeson}, R.~L. and {Chen}, X. and {Ciardi}, D. and {Crane}, M. and {Good}, J. and {Harbut}, M. and {Jackson}, E. and {Kane}, S.~R. and {Laity}, A.~C. and {Leifer}, S. and {Lynn}, M. and {McElroy}, D.~L. and {Papin}, M. and {Plavchan}, P. and {Ram{\'\i}rez}, S.~V. and {Rey}, R. and {von Braun}, K. and {Wittman}, M. and {Abajian}, M. and {Ali}, B. and {Beichman}, C. and {Beekley}, A. and {Berriman}, G.~B. and {Berukoff}, S. and {Bryden}, G. and {Chan}, B. and {Groom}, S. and {Lau}, C. and {Payne}, A.~N. and {Regelson}, M. and {Saucedo}, M. and {Schmitz}, M. and {Stauffer}, J. and {Wyatt}, P. and {Zhang}, A.},
        title = "{The NASA Exoplanet Archive: Data and Tools for Exoplanet Research}",
      journal = {\pasp},
         year = 2013,
        month = aug,
       volume = {125},
       number = {930},
        pages = {989},
          doi = {10.1086/672273},
archivePrefix = {arXiv},
       eprint = {1307.2944},
 primaryClass = {astro-ph.IM},
       adsurl = {https://ui.adsabs.harvard.edu/abs/2013PASP..125..989A}
}

@article{murchikova2020peas,
  title={Peas in a Pod? Radius Correlations in Kepler Multiplanet Systems},
  author={Murchikova, Lena and Tremaine, Scott},
  journal={The Astronomical Journal},
  volume={160},
  number={4},
  pages={160},
  year={2020},
  publisher={The American Astronomical Society}
}

@article{weiss2020kepler,
  title={The Kepler peas in a pod pattern is astrophysical},
  author={Weiss, Lauren M and Petigura, Erik A},
  journal={The Astrophysical Journal Letters},
  volume={893},
  number={1},
  pages={L1},
  year={2020},
  publisher={The American Astronomical Society}
}

@article{he2019architectures,
  title={Architectures of exoplanetary systems--I. A clustered forward model for exoplanetary systems around Kepler’s FGK stars},
  author={He, Matthias Y and Ford, Eric B and Ragozzine, Darin},
  journal={Monthly Notices of the Royal Astronomical Society},
  volume={490},
  number={4},
  pages={4575--4605},
  year={2019},
  publisher={Oxford University Press}
}

@article{lissauer2011architecture,
  title={Architecture and dynamics of Kepler's candidate multiple transiting planet systems},
  author={Lissauer, Jack J and Ragozzine, Darin and Fabrycky, Daniel C and Steffen, Jason H and Ford, Eric B and Jenkins, Jon M and Shporer, Avi and Holman, Matthew J and Rowe, Jason F and Quintana, Elisa V and others},
  journal={The Astrophysical Journal Supplement Series},
  volume={197},
  number={1},
  pages={8},
  year={2011},
  publisher={The American Astronomical Society}
}

@article{fang2012architecture,
  title={Architecture of planetary systems based on Kepler data: Number of planets and coplanarity},
  author={Fang, Julia and Margot, Jean-Luc},
  journal={The Astrophysical Journal},
  volume={761},
  number={2},
  pages={92},
  year={2012},
  publisher={The American Astronomical Society}
}

@article{emsenhuber2023planetary,
  title={Planetary population synthesis and the emergence of four classes of planetary system architectures},
  author={Emsenhuber, Alexandre and Mordasini, Christoph and Burn, Remo},
  journal={The European Physical Journal Plus},
  volume={138},
  number={2},
  pages={181},
  year={2023},
  publisher={Springer}
}

@article{ghosh2024orbital,
  title={Orbital architectures of Kepler multis from dynamical instabilities},
  author={Ghosh, Tuhin and Chatterjee, Sourav},
  journal={Monthly Notices of the Royal Astronomical Society},
  volume={527},
  number={1},
  pages={79--92},
  year={2024},
  publisher={Oxford University Press}
}

@article{winn2015occurrence,
  title={The occurrence and architecture of exoplanetary systems},
  author={Winn, Joshua N and Fabrycky, Daniel C},
  journal={Annual Review of Astronomy and Astrophysics},
  volume={53},
  number={1},
  pages={409--447},
  year={2015},
  publisher={Annual Reviews}
}

@article{smith2009orbital,
  title={Orbital stability of systems of closely-spaced planets},
  author={Smith, Andrew W and Lissauer, Jack J},
  journal={Icarus},
  volume={201},
  number={1},
  pages={381--394},
  year={2009},
  publisher={Elsevier}
}

@article{wilson1927probable,
  title={Probable inference, the law of succession, and statistical inference},
  author={Wilson, Edwin B},
  journal={Journal of the American Statistical Association},
  volume={22},
  number={158},
  pages={209--212},
  year={1927},
  publisher={Taylor \& Francis}
}

@article{duffell2015eccentric,
  title={ECCENTRIC JUPITERS VIA DISK--PLANET INTERACTIONS},
  author={Duffell, Paul C and Chiang, Eugene},
  journal={The Astrophysical Journal},
  volume={812},
  number={2},
  pages={94},
  year={2015},
  publisher={The American Astronomical Society}
}

@article{chambers1998making,
  title={Making the terrestrial planets: N-body integrations of planetary embryos in three dimensions},
  author={Chambers, JE and Wetherill, GW},
  journal={Icarus},
  volume={136},
  number={2},
  pages={304--327},
  year={1998},
  publisher={Elsevier}
}

@article{muller2024mass,
  title={The mass-radius relation of exoplanets revisited},
  author={M{\"u}ller, Simon and Baron, Jana and Helled, Ravit and Bouchy, Fran{\c{c}}ois and Parc, L{\'e}na},
  journal={Astronomy \& Astrophysics},
  volume={686},
  pages={A296},
  year={2024},
  publisher={EDP Sciences}
}

@book{edgington2007randomization,
  title={Randomization tests},
  author={Edgington, Eugene and Onghena, Patrick},
  year={2007},
  publisher={CRC press}
}

@article{lambrechts2014separating,
  title={Separating gas-giant and ice-giant planets by halting pebble accretion},
  author={Lambrechts, Michiel and Johansen, Anders and Morbidelli, Alessandro},
  journal={Astronomy \& Astrophysics},
  volume={572},
  pages={A35},
  year={2014},
  publisher={EDP Sciences}
}

@article{bitsch2015growth,
  title={The growth of planets by pebble accretion in evolving protoplanetary discs},
  author={Bitsch, Bertram and Lambrechts, Michiel and Johansen, Anders},
  journal={Astronomy \& Astrophysics},
  volume={582},
  pages={A112},
  year={2015},
  publisher={EDP Sciences}
}

@article{mann2009circumstellar,
  title={The circumstellar disk mass distribution in the Orion Trapezium cluster},
  author={Mann, Rita K and Williams, Jonathan P},
  journal={The Astrophysical Journal},
  volume={694},
  number={1},
  pages={L36--L40},
  year={2009},
  publisher={The American Astronomical Society}
}

@article{pascucci2016steeper,
  title={A steeper than linear disk mass--stellar mass scaling relation},
  author={Pascucci, Ilaria and Testi, Leonardo and Herczeg, Gregory J and Long, F and Manara, CF and Hendler, N and Mulders, Gijs D and Krijt, S and Ciesla, F and Henning, Th and others},
  journal={The Astrophysical Journal},
  volume={831},
  number={2},
  pages={125},
  year={2016},
  publisher={The American Astronomical Society}
}

@article{he2022debiasing,
  title={Debiasing the Minimum-mass Extrasolar Nebula: On the Diversity of Solid Disk Profiles},
  author={He, Matthias Y and Ford, Eric B},
  journal={The Astronomical Journal},
  volume={164},
  number={5},
  pages={210},
  year={2022},
  publisher={The American Astronomical Society}
}
\bibliographystyle{aasjournalv7}

\end{document}